\documentclass[%
 aip,amsmath,amssymb,
 reprint,%
jcp]{revtex4-1}
\usepackage{amsmath} 
\usepackage{amsthm} 
\usepackage{amssymb}	
\usepackage{graphicx} 
\usepackage{hyperref}
\makeatletter 
\renewcommand{\v}[1]{\ensuremath{\mathbf{#1}}} 
 
\newcommand{\matrixel}[3]{\left< #1 \vphantom{#2#3} \right|
 #2 \left| #3 \vphantom{#1#2} \right>} 
\newcommand*{\redefinesymbolwitharg}[1]{%
  \expandafter\newcommand\csname ltx#1\endcsname{}%
  \expandafter\let\csname ltx#1\expandafter\endcsname\csname #1\endcsname
  \expandafter\renewcommand\csname #1\endcsname[1]{%
   \csname ltx#1\endcsname\left(##1\right)%
 }%
}
\let\baraccent=\= 
\renewcommand{\=}[1]{\stackrel{#1}{=}} 

\theoremstyle{definition}

\theoremstyle{remark}

\redefinesymbolwitharg{sin}
\redefinesymbolwitharg{cos}
\redefinesymbolwitharg{tan}
\redefinesymbolwitharg{cot}
\redefinesymbolwitharg{sec}
\redefinesymbolwitharg{csc}

\redefinesymbolwitharg{sinh}
\redefinesymbolwitharg{cosh}
\redefinesymbolwitharg{tanh}
\redefinesymbolwitharg{coth}

\redefinesymbolwitharg{arcsin}
\redefinesymbolwitharg{arccos}
\redefinesymbolwitharg{arctan}

\redefinesymbolwitharg{exp}
\redefinesymbolwitharg{log}
\redefinesymbolwitharg{ln}

\usepackage{graphicx}
\usepackage{dcolumn}
\usepackage{bm}
\usepackage{braket}
\usepackage{mathrsfs}
\usepackage{color}
\begin{document}
\preprint{AIP/123-QED}

\title[Experimentally Attainable Optimal Pulse Shapes Obtained with the Aid of Genetic Algorithms]{Experimentally Attainable Optimal Pulse Shapes Obtained with the Aid of Genetic Algorithms}

\author{Rub{\'e}n D. Guerrero}
\affiliation{
Department of Physics, Universidad Nacional de Colombia.}
 \email{rdguerrerom@unal.edu.co}

\author{Carlos A. Arango}%
\affiliation{ Departament of Chemical Sciences, Universidad Icesi, Cali, Colombia.}
\email{caarango@icesi.edu.co}

\author{Andr{\'e}s Reyes}
\affiliation{%
Department of Chemistry, Universidad Nacional de Colombia.}%
 \email{areyesv@unal.edu.co}
\date{\today}
\setlength{\parindent}{0pt}

\begin{abstract}
We propose a methodology to design optimal pulses for achieving quantum optimal control on molecular systems. Our approach constrains pulse shapes to linear combinations of a fixed number of experimentally relevant pulse functions. Quantum optimal control is obtained by maximizing a multi-target fitness function with genetic algorithms. As a first application of the methodology we generated an optimal pulse that successfully maximized the yield on a selected dissociation channel of a diatomic molecule. Our pulse is obtained as a linear combination of linearly chirped pulse functions. Data recorded along the evolution of the genetic algorithm contained important information regarding the interplay between radiative and diabatic processes. We performed a principal component analysis on these data to retrieve the most relevant processes along the optimal path. Our proposed methodology could be useful for  performing quantum optimal control on more complex systems by employing a wider variety of pulse shape functions. 
\end{abstract}

\maketitle

\section{\label{sec:Intro}Introduction}
Quantum optimal control (QOC) methodologies are employed to produce optimal pulses capable of steering the quantum state of a molecule towards a given target state. QOC is accomplished by using the matter-radiation interaction and the time-frequency coherence of light.\cite{warren1993coherent, shapiro2003principles}
\\

Among all theoretical QOC approaches, gradient-based feedback (GBF) methods are widely employed due to their robustness and rapid convergence.\cite{zhu1998rapidly} Although GBF approaches have been used to compute pulses capable of attaining desired target molecular states,\cite{werschnik2007quantum, guerrero2014optimal,kormann2010fourier} their applicability still faces serious limitations. On one hand, the convergence of GBF approaches towards the closest optimum is biased by the choice of the initial guess for the field. 
On the other hand, the precise reproduction in the laboratory of the generated optimal pulses  is still challenging due to their complexity.
\\

In the past two decades, alternative QOC approaches based on heuristic optimization strategies have been proposed to improve the convergence towards the global optimum. Among these heuristic schemes, Genetic Algorithms (GA) have become the methods of choice. \cite{ amstrup1995identification, michalewicz1996genetic, phan1999self, tesch2001applying, brixner2001problem, fogel2006evolutionary, tsubouchi2008rovibrational}
For instance, optimal pulses in the frequency domain have been generated with the aid of GA to control molecular events relevant for quantum information processing.\cite{tesch2001applying, tsubouchi2008rovibrational}
In spite of overcoming the limitation of GBF methods to reach the global optimum, current GA implementations still fail at generating optimal pulses that can be precisely reproduced in the laboratory.
\\

A desirable QOC approach designed to overcome the aforementioned limitations must guarantee the convergence towards the global optimum and constrain the pulse generation to linear combinations of experimentally attainable pulse functions.
\\

We propose a novel QOC method designed to overcome these limitations. Our approach introduces a GA methodology to optimize the parameters of a linear combination of analytical pulse shapes, which are experimentally synthetizable with linear optics techniques. As a first test of the efficiency of the method, we control the dissociation yield of a diatomic molecule through a selected channel. 
In addition to controlling molecular processes, the proposed methodology generates data that can be used to analyze the interplay between diabatic and photodynamic processes along the optimal path. To this aim we perform a statistical principal component analysis (PCA) on the distribution of time integrated transition moment integrals calculated during the GA evolution.
The resulting PCA singular values and vectors provide key information about the intricate sequence of events driven by the optimal pulse.\\

The remainder of this paper is organized as follows: section \ref{sec:Methods} introduces the notation, the theoretical background and the proposed GA methodology; section \ref{sec:Apps} shows the application of this methodology to perform optimal control of the dissociation yields of a model diatomic molecule; section \ref{sec:Conclusions} provides some concluding remarks and perspectives of this work. 

\section{\label{sec:Methods}Theory}
The wavefunction, $\Psi$, of a molecular system composed of $N$ nuclei with coordinates $\v{R}=(\v{R_1},\v{R_2},\dots, \v{R_I}, \dots, \v{R_N})$, and $n$ electrons with coordinates $\v{r}=(\v{r_1},\v{r_2},\dots,\v{r_i}, \dots,\v{r_n})$ can be calculated by solving the time-dependent Schr\"odinger equation 
(atomic units will be used hereafter)
\begin{align}\label{eq:Schro}
\frac{\partial \Psi}{\partial t}=-i {\hat H} \Psi,
\end{align}
here ${\hat H}=\mathcal{\hat H}_{0}+\mathcal{\hat H}^{\mathrm{int}}$, where $\mathcal{\hat H}_{0}$ is the molecular Hamiltonian, and $\mathcal{\hat H}^{\mathrm{int}}$ is the interaction of the system with the pulse.
\\

Under the Born-Oppenheimer ansatz\cite{doltsinis2002first} for the  wavefunction of Eq. \ref{eq:Schro}, the evolution of the nuclear wavepacket, $\chi_k$, can be obtained by tracing out the electronic degrees of freedom:
\begin{subequations}
\label{eq:adiabatic}
\begin{equation}
\left[\mathcal{T}(\v{R}) +  E_{k}(\v{R}) \right]\chi_{k}(\v{R};t) + \sum_{l}{\mathcal{C}_{kl}}\chi_{k}(\v{R};t)=i\frac{\partial}{\partial t} \chi_{k}(\v{R};t),
\end{equation}
\begin{equation}
{\mathcal{C}_{kl}}(\v{R})=\matrixel{\Phi_{k}}{\mathcal{T}(\v{R})}{\Phi_{k}} -\sum_{J}\frac{1}{M_{J}}\matrixel{\Phi_{k}}{\v{\nabla}_{J}}{\Phi_{l}}\v{\nabla}_{J},
\end{equation}
\end{subequations}
here $E_{k}(\v{R})$ is the potential energy surface (PES) corresponding to electronic state $\ket{\Phi_k}$, and ${\mathcal{C}_{kl}}(\v{R})$ are diabatic couplings between electronic states $\ket{\Phi_k}$ and $\ket{\Phi_l}$. 
\\

The radiation-matter interaction, $\mathcal{\hat H}^{\mathrm{int}}$, is considered semiclassically within the dipolar approximation, 
\begin{align}\label{eq:min_coup}
\mathcal{H}^{\mathrm{int}}_{_{kl}}(\v{R},t)&= -4\pi \mu_{kl}(R)\varepsilon(t),\\
\varepsilon(t) &= \sum_{i}\mathcal{V}_{i}(t), \nonumber
\end{align}
where $\mu_{kl}$ is an element of the electronic transition dipole matrix.\\

We propose to express the pulse, $\varepsilon(t)$, as a linear combination of linearly chirped pulse (LCP) functions. These LCPs have been used successfully to control photochemical reactions.\cite{ruhman1990application, szakacs1994locking, zhang1998wavepacket, carini2012quantum, huang2012formation, amaran2013femtosecond, carini2013production, huang2014creation}
The time profile of a LCP can be written as:
\begin{align}\label{eq:Chirp}
\mathcal{V}(t)=E_{0}\exp{-\frac{(t-\tau_0)^2}{2\tau^2}}\cos{\frac{1}{2}c(t-\tau_0)^2+\omega_{0}(t-\tau_0)},
\end{align}
here $\tau_0$ is the time shift, $E_0$ is the pulse amplitude, $c$ is the chirp rate, $\omega_0$ is the central frequency, and $\tau$ is the pulse width. As observed, the instantaneous frequency of a LCP changes linearly with time.
\\

The nuclear state of the molecule, $\ket{X(\v{R};t)}$, can be expressed in a diabatic basis as a vector of nuclear configurations:
\begin{align}
\ket{X(\v{R};t)}=\begin{pmatrix}
\chi_{1}(\v{R};t)\\
\vdots\\
\chi_{N}(\v{R};t)
\end{pmatrix}.
\end{align}
The formal solution of Eq. \ref{eq:Schro} is the Green's function of the system of coupled equations of Eq. \ref{eq:adiabatic}:
\begin{align}
\ket{X(\v{R};t+2\delta t)}=e^{-i\mathcal{H}2\delta t}\ket{X(\v{R};t)}.
\end{align}
This Green's function can be approximated \footnote{Because of the coupling to the field, ${\hat U}= e^{-i\int_{0}^{2\delta t} [\mathcal{\hat H}(t)]dt 2 \delta t} \approx e^{-i\mathcal{\hat H}(\delta t) 2 \delta t^2}$. This is a good approximation if $\omega \delta t \ll 1$, with $\hbar \omega$ as a typical energy spacing at the minimum energy of the processes taking place in the simulation. For our setup, $\delta t =1.875\times 10^{-1}$ and $\omega \approx 9\times 10^{-3}$} 
in terms of a symmetric double Trotter expansion to obtain: 
\begin{align}\label{eq:Trotter}
e^{-i\mathcal{H}2\delta t} &\approx e^{-i\mathcal{T}\delta t} e^{-i\mathcal{V}2\delta t}  e^{-i\mathcal{T}\delta t}.
\end{align}
The propagation of the initial (diabatic) nuclear wavepacket for each $\delta t$, is achieved by switching back and forth between the diabatic and adiabatic representations of the wavepacket.\cite{alvarellos1988evolution}

\subsection{Genetic Algorithm}

The GA scores the individuals using a fitness function, $J_{\hat O}$. For an observable ${\hat O}$ the fitness function is 
\begin{align}
J_{\hat O} &= J^{(1)}_{\hat O}[\Psi]+ J^{(2)}[\varepsilon(t)],\\
J^{(1)}_{\hat O}[\Psi] &=\matrixel{\Psi(T)}{\hat O }{\Psi(T)}, \\
J^{(2)}[\varepsilon(t)] &= -\int_{0}^{T}\|\varepsilon(t)\|^{2}dt
\end{align}
here $J^{(2)}[\varepsilon(t)]$ penalizes the fluence of the laser. We propose the multi-target fitness function as:  
\begin{align}\label{eq:fitness}
J &= J_{{\hat O}_1}+\sum_{i=2}^{k}\frac{1}{J_{{\hat O}_i}}
\end{align}
to maximize observable ${\hat O}_1$, while minimizing observables ${\hat O}_{i\ne 1}$.
\\ 

In our GA implementation, the chromosome of the $i$-th individual is built by concatenating the vectors of genes of the $N$ LCPs, $\v{E}_{i}=\{y_{1},y_{2},\cdots,y_{N}\}$. Each LCP is represented by a 5-vector $y_{j}=(E_0, \tau_0, c, \tau, \omega_0)$. 
We propose two genetic operations to evolve the $N$ LCPs: (1) a mutation to change one or more genes with probability $\boldsymbol{\pi}_{M}$, and (2) a crossover operation acting on two individuals to generate a new individual with probability $\boldsymbol{\pi}_{X}$. 
The new chromosome is generated as a fitness-weighted linear combination of the parents. \\

Optimizations were carried out following the next steps:
\begin{enumerate}
\item[0.] Initialize the population with a list of individuals whose chromosomes were generated randomly. \footnote{It is crucial to limit the generation of the initial list of individuals to the region of the parameters space with experimental relevance.}
\item[1.]{Propagate individuals using the same initial condition and calculate their scores with the fitness function of Eq. \ref{eq:fitness}.}
\item[2.]{Sort individuals in descending order according to their fitness.}
\item[3.]{Compute the cumulative fitness of the entire population.}
\item[4.]{Keep the individuals with fitness greater or equal than the cumulative fitness. Discard the remainder.}
\item[5.]{Replace those individuals discarded in step 4 with new ones generated by crossing over the survivors with probability $\boldsymbol{\pi}_{X}$.}
\item[6.]{Go through the new population list sampling the probability of mutation of each gene with a Monte Carlo scheme. Mutate those genes with probability lower than $\pi_M$}.
\item[7.]{Go back to step 1 if the maximum number of generations has not been reached.}
\end{enumerate}

\section{\label{sec:Apps} NUMERICAL TEST}
As a first application of the proposed methodology we selectively controlled the yields through the dissociation channels of a model diatomic molecule.\cite{gross1992optimal}
This model could be related to the control over the yield of homolytic or heterolytic bond cleavage.\cite{warren1993coherent} 
\\

\begin{figure}[htp]
\caption{Time dependence of the populations (panel A) and currents (panel B) on the dissociative channels $E_{2}(R)$ and $E_{3}(R)$ for the free evolution of the wavepacket. Results for $E_{1}(R)$ are not reported, as this curve is neither coupled to $E_{2}(R)$ nor $E_{3}(R)$.}
\includegraphics[]{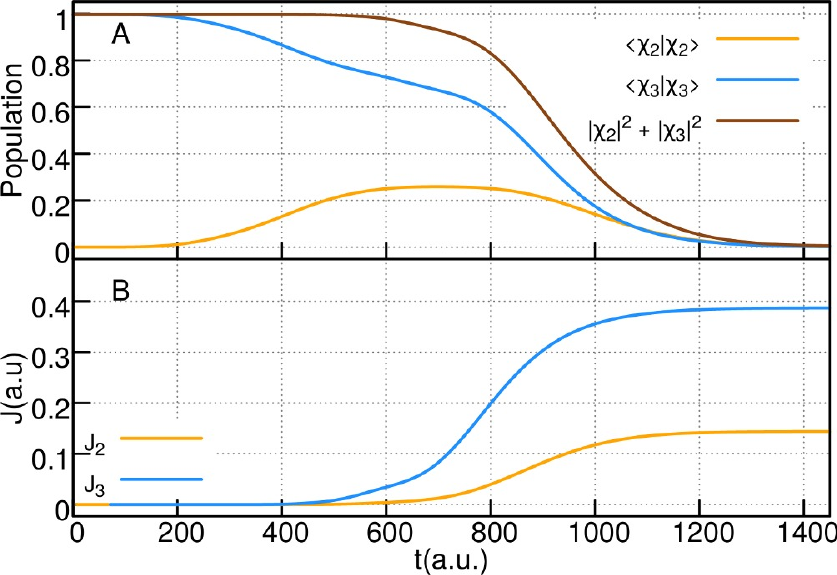}\label{fig:Free_results}
\end{figure}

The molecule is modeled with three PESs: $E_1(R)$, $E_2(R)$, and $E_3(R)$ which are coupled diabatically.\cite{gross1992optimal} 
The off-diagonal terms of the nuclear Hamiltonian matrix are 
\begin{align}
&\mathcal{H}^{\mathrm{int}}_{12}=\mu_{12}(R)\varepsilon(t),\\
&\mathcal{H}^{\mathrm{int}}_{13}=\mu_{13}(R)\varepsilon(t),\\
&H_{23}= \mathcal{H}^{\mathrm{int}}_{_{23}}+\mathcal{C}_{23}(R) =\mu_{23}(R)\varepsilon(t)+\mathcal{C}_{23}(R).
\end{align}
Table \ref{tab:Hamiltonian} lists all the terms and parameters employed in the time propagation of the molecular wavepacket. 
\\
\begin{table}[h]
\caption{Parameters defining the PESs and diabatic couplings. Time and space discretization values are provided. \label{tab:Hamiltonian}}
\begin{tabular}{l}
\hline\hline
Ground state potential\\
\hline
$E_{1}(R)=D_e(1-\exp{-y(R-r_e)})^2$\\
$D_e=0.075$\\
$\omega_e=0.0077782$ \\
$r_e=4.125$\\
$M=1836.0$\\
$y=\omega_e\sqrt{M/2/D_e}$\\
\hline
Dissociative potentials\\
\hline
$E_{\alpha}(R)= m_{\alpha}(R-3.15)+b_{\alpha}+c \exp{-d(R-3.15)}$\\
\begin{tabular}{ll}
$m_2=-0.008681$    & $m_3=-0.01736$\\
$b_2=0.1805$    & $b_3=0.2$\\
$c=0.25$ & \\
$d=5.0$ & \\
\end{tabular}\\

\hline
Dipole functions and diabatic couplings\\
\hline
$\mu_{12}(R)=\mu_{23}(R)=\mu \exp{-a(R-r_x)^2}$\\
\begin{tabular}{ll}
$\mu_{13}(R) = \mu \tanh{-10.0(8.0-R)}$ & $r> r_x$\\
$\mu_{13}(R) = \mu \tanh{-10.0(r-3.0)}$ & $r< r_x$\\
$\mu=0.2$;& \\
$a=5.0$& \\
$r_x=5.25$& \\
$\mathcal{C}_{23}(R)=A  \exp{-a(R-r_x)^2}$ & \\
\end{tabular}\\
\begin{tabular}{ll}
Intermediate coupling:    & $A=0.003$\\ 
Strong coupling:    & $A=0.0075$\\
\end{tabular}\\
\hline
Optical potentials \cite{gross1992optimal}\\
\hline
Grid spacing $\Delta r =4.6875\times 10^{-2}$\\ 
Number of grid points $N_r=220$ \\
Time step $\Delta t=1.875\times 10^{-1}$ \\
Propagation steps $N_t=8192$\\
\hline
\end{tabular}
\end{table}
We initialize the propagation by placing a Gaussian wavepacket on the PES $E_{3}(R)$, with center at $R=r_e$ and width $0.265$. The free evolution of this wavepacket is displayed in Figure \ref{fig:Free_results}. \footnote{Diabatic representation of the dynamics was used through the paper} Panel A shows the population on the $i$-th PES, $\left<\chi_i\mid\chi_i\right>$, while panel B displays the currents $J_2$ and $J_3$, calculated using 
\begin{align}\label{eq:J_target_1}
J_{i}&=\frac{1}{M}\int_{0}^{T}\mathfrak{I}\matrixel{\chi_{i}}{\hat p}{\chi_{i}}\delta(R-R_d)dt.
\end{align}
Here $\hat{p}$ is the momentum operator, $R_d$ is the dissociation limit distance, and $T$ is the final time of the propagation. As observed in Figure \ref{fig:Free_results}, the norm is not conserved along the propagation since we imposed absorbing boundary conditions by employing optical potentials. An interesting feature of the free evolution of the population shown in panel A is that the diabatic coupling induced population transfer between the PESs for times $t>200$.
\\

Our QOC goal was to maximize the nuclear probability flux $J_2$ while minimizing the flux $J_3$ beyond the dissociative limit $R_d=8.4$. To reach this aim we employed the following fitness function:
\begin{align}\label{eq:J_target}
J= J_{2}+\frac{1}{J_{3}}.
\end{align}
\begin{table}[h]
\caption{\label{tab:boundaries} Boundaries of the LCP parameters space used in our GA approach.}
   \begin{tabular}{|ccc|}
    \hline
    Parameter & Minimum & Maximum \\ \hline
    $E_0$         & 0.01     & 0.2    \\
    $\tau_0$        &   $1.2\times 10^{2}$   & $9.0\times 10^{2}$    \\
    $\tau$         & 20.0    &    60.0\\
    $\omega$         & 0.14     & 0.16    \\
    $c$         &  $-5.0\times 10^{-7}$    & $5.0\times 10^{-7}$    \\ \hline
    \end{tabular}
\end{table}

\begin{figure}[htp]
\caption{Optimal pulses calculated to maximize $J$ (Eq. \ref{eq:J_target}). Panels A, B and C plot pulses assembled with 30, 40, and 60 LCPs, respectively.}
\includegraphics[]{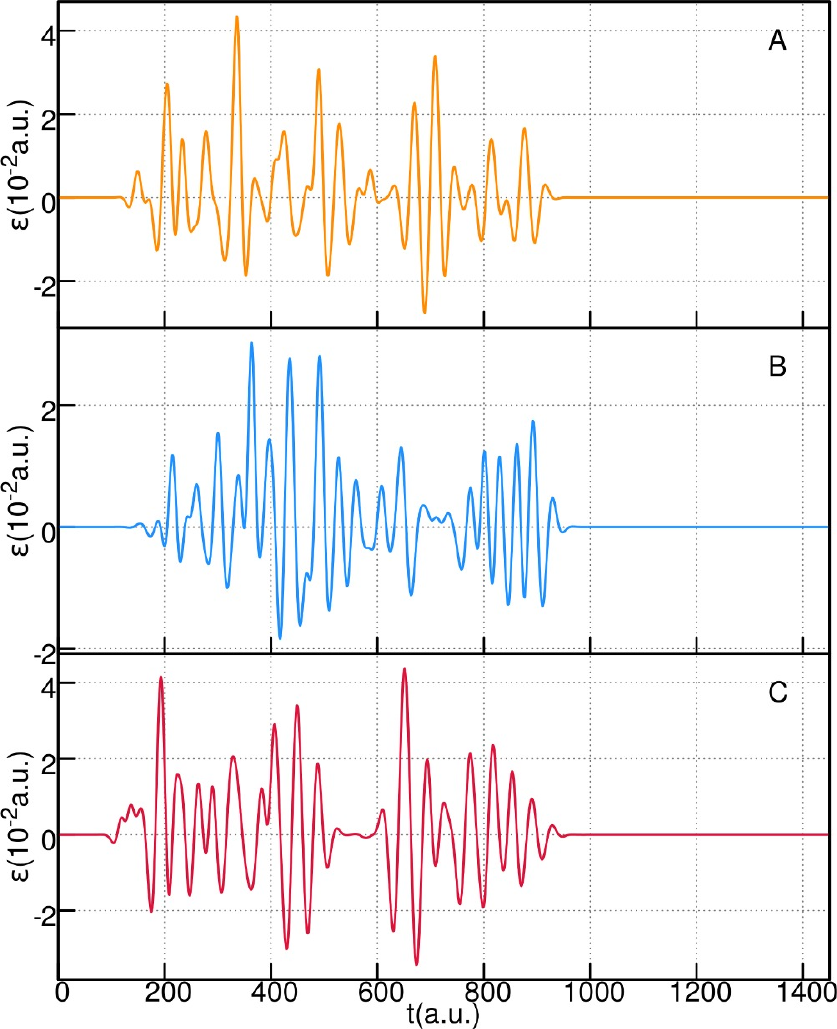}\label{fig:Pulses}
\end{figure}

\begin{figure}[htp]
\caption{Power spectra of the optimal pulses plotted in Figure \ref{fig:Pulses}.}
\includegraphics[]{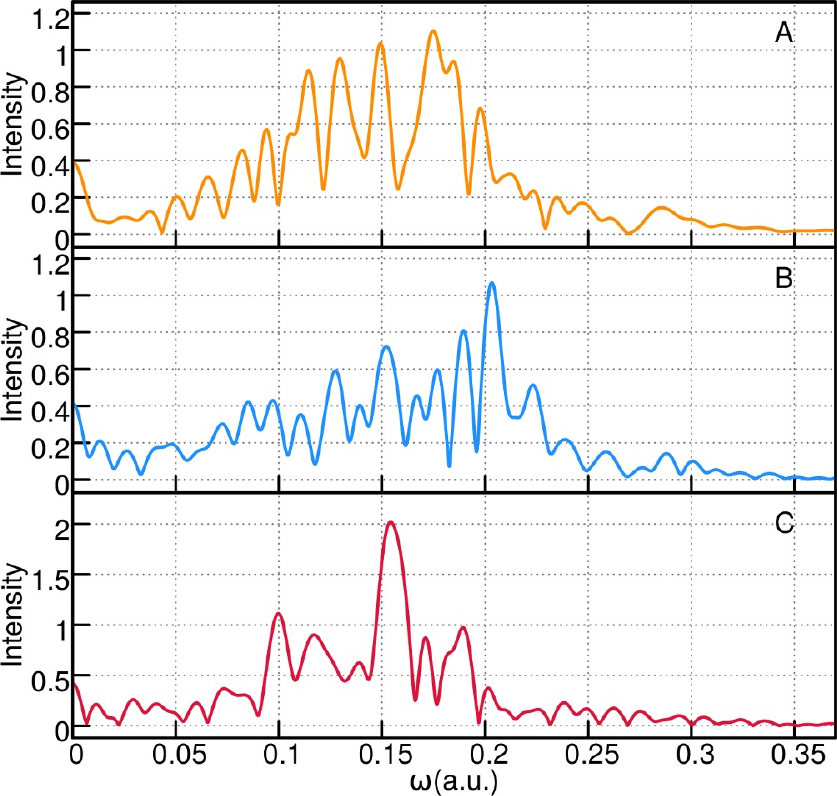}\label{fig:Spectrum}
\end{figure}

We analyzed the dependence of the number of LCPs, used in the GA optimization, on the effectiveness of the pulse in driving the system towards the selected target. To this aim, we tested three pulses built with 30, 40, and 60 LCPs. These pulses are shown in time and frequency domains in Figures \ref{fig:Pulses} and \ref{fig:Spectrum}. Figure \ref{fig:Spectrum} reveals that the spectra of these pulses is band-width limited; this is a desirable feature for tailoring pulses in the laboratory. The spectrum shown in panel C of Figure \ref{fig:Spectrum} displays a dominant band around $\omega_0=0.154$, frequency that is in the range of the transitions $3\rightarrow1$ and $1\rightarrow2$.  

\begin{figure}[h!t!]
\caption{The free evolution of the wavepacket on curves $E_{3}(R)$, $E_{2}(R)$, and $E_{1}(R)$ is shown on panels A, B, and C, respectively. 
Panels D, E, and F show the time dependence of the wavepacket driven by the optimal pulse plotted on Figure \ref{fig:Pulses}C on curves $E_{3}(R)$, $E_{2}(R)$, and $E_{1}(R)$,  respectively.} 
\includegraphics[]{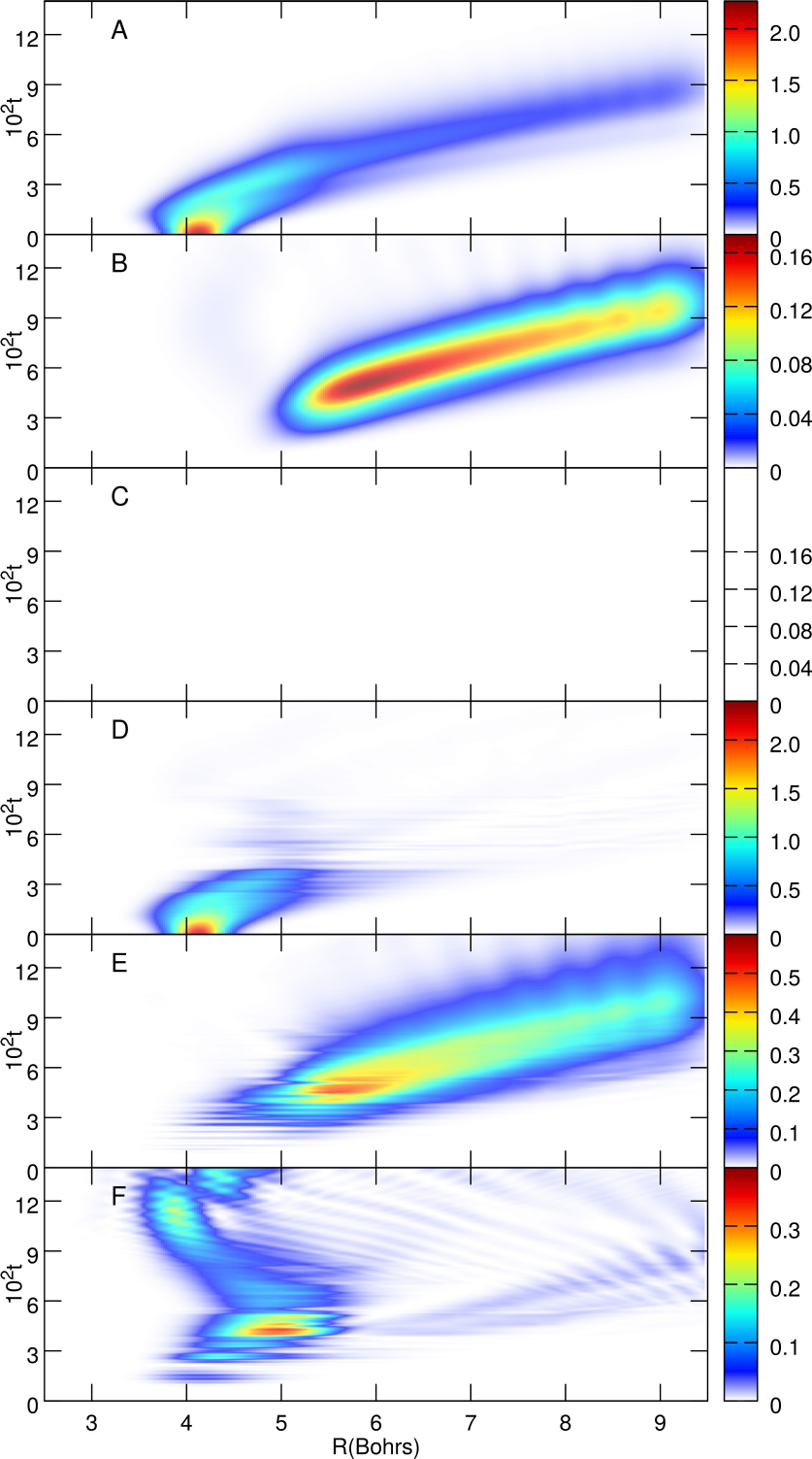}\label{fig:Wavepacket}
\end{figure}

\begin{figure}[h!]
\caption{Time-integrated currents as function of the number of LCPs. Yellow, blue, and red lines represent the currents obtained with optimal pulses built with 30, 40 and 60 LCPs,  respectively.}
\includegraphics[]{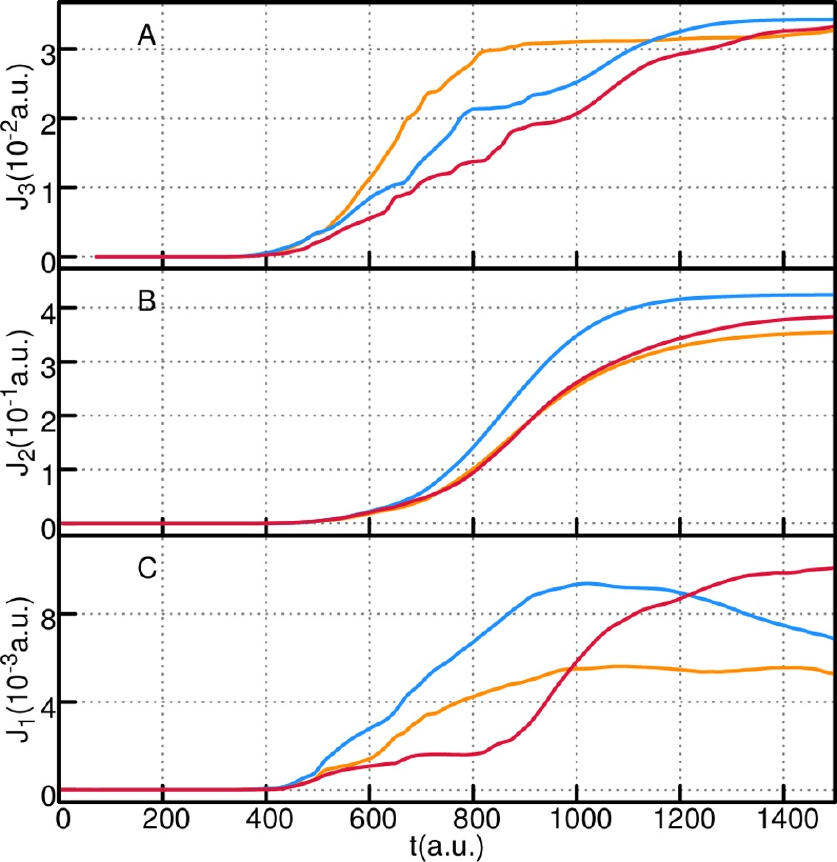}\label{fig:Targets}
\end{figure}

Panels D, E, and F in Figure \ref{fig:Wavepacket} present the evolution of the wavepacket driven by the pulse on the PES $E_{3}(R)$, $E_{2}(R)$, and $E_{1}(R)$, respectively. 
As shown in Figure \ref{fig:Pulses}C, the pulse displays an intense peak at $t\approx 200$, which drives a significant portion of the initial wavepacket towards the electronic ground state. This process is shown in more detail in panels D and F of Figure \ref{fig:Wavepacket}. Simultaneously, the pulse induced transitions from $E_1(R)$ to $E_{2}(R)$ and $E_{3}(R)$. 
\\
We now analyze the efficiency of each of the pulses of Figure \ref{fig:Pulses} in attaining the proposed target. We perform this analysis in terms of time-integrated probability currents $J_i$ of Eq. \eqref{eq:J_target_1}. 
Panels A, B, and C of Figure \ref{fig:Targets} display the integrated currents at the dissociation limit ($R_d=8.4$) on channels $E_{1}(R)$, $E_{2}(R)$, and $E_{3}(R)$, 
respectively. As observed, each pulse steers the system successfully towards the proposed goal: maximizing $J_2$ while minimizing $J_3$. We employed the ratio $J_2/J_3$ at time 1400 to quantify the efficiency of each pulse to reach the proposed goal. The resulting ratios $J_2/J_3=10.9$, $11.4$, $12.3$ for pulses A, B, and C, respectively, suggest that the efficiency of the optimal pulse increases with the number of LCPs.

\begin{figure}[htp]
\caption{Time evolution of the populations on the PESs driven by the optimal pulse of Figure \ref{fig:Pulses}C.}
\includegraphics[]{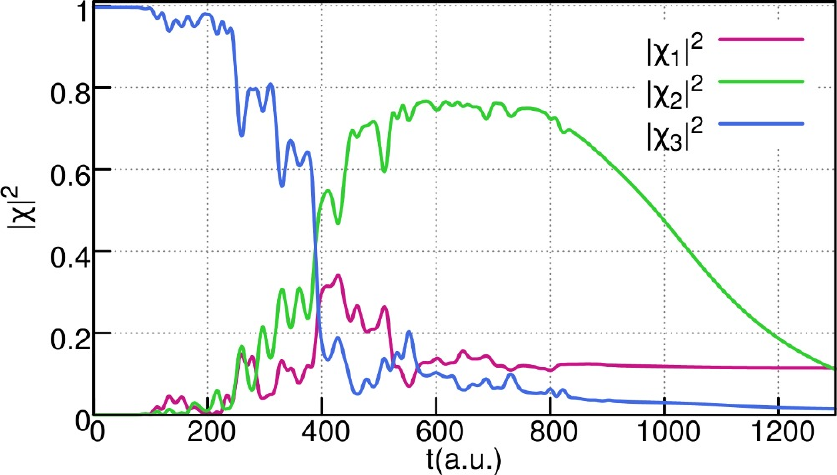}\label{fig:Population_opt}
\end{figure}

The evolution of the populations induced by the optimal pulse of Figure \ref{fig:Pulses}C is reported in Figure \ref{fig:Population_opt}. As observed, at $t\approx 400$ the pulse transferred approximately half of the wavepacket from $E_{3}(R)$ to $E_{1}(R)$ and $E_{2}(R)$. In contrast, Figure \ref{fig:Free_results}A shows that the free evolution of the populations at $t=400$ was barely affected by the diabatic coupling. 
\\

\subsection{Principal component analysis}
With the aid of statistical tools we can enrich our understanding of the intricate sequence of events resulting from the effects of the pulse and the diabatic coupling on the wavepacket. 

To this aim, we first built a Markov chain by recording the time-integrated transition moment integrals (TI-TMI) of each surviving individual along the GA evolution. Each TI-TMI is constructed as the vector $(P_{1\rightarrow 2},P_{2\rightarrow 3},P_{3\rightarrow 1})$, where:
\begin{align}
P_{i\rightarrow j} = \int_{0}^{T}|\matrixel{\chi_{i}(t)}{ R}{\chi_{j}(t)}|^2 dt/T.
\end{align}
The resulting distribution of the TI-TMIs narrows around the most frequent optical transitions along the optimal path. \\

A principal component analysis (PCA) performed on this Markov chain provides singular values and vectors that can be used to enhance our understanding of the system. For instance, it supplies information about the main transitions induced by the pulse. 
Table \ref{tab:PCA} summarizes the PCA results for the pulse of Figure \ref{fig:Pulses}C when acting on the molecule. The absolute value of the components of each singular vector measures the contribution of the transitions occurring along the optimal path.

\begin{table}[h! b!]
\caption{\label{tab:PCA} PCA results for the photo-processes induced by the pulse of Figure \ref{fig:Pulses}C. $P_{i\rightarrow j}$ is the  transition amplitude between PESs $i$ and $j$ and $W_{312}$ is the singular value for the corresponding process.}
\begin{tabular}{|lcc|}
\hline
     & \textbf{Process 1} & \textbf{Process 2}  \\ \hline
\textbf{$P_{1\rightarrow 2}$} & -0.150           &    0.825                \\ 
\textbf{$P_{2\rightarrow 3}$} & 0.770           &    -0.253                 \\ 
\textbf{$P_{3\rightarrow 1}$} & -0.620            &    -0.500               \\ 
\textbf{$W_{312}$}            & 1.126 & 1.100  \\ \hline
\end{tabular}
\end{table}

Table \ref{tab:PCA} lists the 
singular values, $W_{312}$, that quantify the contribution of the process along the optimal path. The time-integrated probability amplitudes, $P_{i\rightarrow j}$, for each optical transition are also reported in this table.  
\\
We obtained two dominant processes: process 1 with singular value $1.126$, and process 2 with singular value $1.100$. As observed in Table \ref{tab:PCA}, process 1 comprises mainly optical transitions $P_{3\rightarrow 2}$ and $P_{3\rightarrow 1}$. 
This process is most likely associated with the earlier stages of the dynamics ($t<500$) where a significant portion of the initial wavepacket was driven towards the ground PES. This could be seen in  Figure \ref{fig:Pulses}C. Process 2 comprises mainly optical transitions $P_{3\rightarrow 1}$ and $P_{1 \rightarrow 2}$. This process is most likely related with the dynamics for $t>500$, where the pulse drives the portion of the wavepacket in the ground state towards its dissociation through PESs $E_{2}$ and $E_{3}$. This could be seen in Figure \ref{fig:Pulses}C.

\section{\label{sec:Conclusions}Conclusions} 

The methodology proposed in this work achieves QOC employing GA and constraining the optimal pulse to analytical functions. The resulting pulse can be reproduced in the laboratory with the current technology.  
\\
The successful application of this QOC approach relies on the choice of a basis set of pulse functions. These functions must have spectral properties compatible with the transition energies of the system. 
\\
The proposed methodology is robust and general, thereby allowing the use of linear superpositions of other experimentally available pulse functions. The suitable selection of the basis set of pulse functions will improve the performance and accuracy of the method to achieve more complex targets. 
\\
The PCA performed on the Markov chain of the TI-TMIs shed light on the photo-dynamical processes induced by the optimal pulse. It proved to be an excellent tool for obtaining quantitative information about the optical processes taking place along the optimal path. 
\\
We will apply our approach to control more complex systems. For instance, we will optimize sequences of pulses in Coherent Anti-Stokes Raman Scattering (CARS),\cite{begley1974coherent,malinovskaya2007chirped} to control the dynamics in processes such as proton-transfer and  photoisomerization.\\

\noindent{\small \bf ACKNOWLEDGMENTS}\\
RDG thanks COLCIENCIAS and DIB-UNAL for their financial support. CAA thanks Universidad Icesi for its financial support. The authors are grateful to Jorge Ali, Jonathan Romero, Johan Galindo, Giovanny Rojas and T. J. Martinez for their detailed proofreading and helpful comments of this paper.
\bibliographystyle{unsrt}

\end{document}